# Broad Band Optical Properties of Large Area Monolayer CVD Molybdenum Disulfide

Wei Li[1,2], A. Glen Birdwell[3], Matin Amani[3], Robert A. Burke[3], Xi Ling[4], Yi-Hsien Lee[4,5], Xuelei Liang[2], Lianmao Peng[2], Curt A. Richter[1], Jing Kong[4], David J. Gundlach[1], and N. V. Nguyen[1]

1. Semiconductor and Dimensional Metrology Division, National Institute of Standards and Technology, Gaithersburg, MD 20899, USA
2. Key Laboratory for the Physics and Chemistry of Nanodevices and Department of Electronics, Peking University, Beijing 100871, China
3. Sensors and Electron Devices Directorate, US Army Research Laboratory, Adelphi, Maryland 20783, USA
4. Department of Electrical Engineering and Computer Science, Massachusetts Institute of Technology, Cambridge, MA 02139, USA
5. Material Sciences and Engineering, National Tsing-Hua University, Hsinchu, 30013, Taiwan

Corresponding authors: nhan.nguyen@nist.gov, liangxl@pku.edu.cn.

ABSTRACT: Recently emerging large-area single-layer $MoS_2$ grown by chemical vapor deposition has triggered great interest due to its exciting potential for applications in advanced electronic and optoelectronic devices. Unlike gapless graphene, $MoS_2$ has an intrinsic band gap in the visible which crosses over from an indirect to a direct gap when reduced to a single atomic layer. In this article, we report a comprehensive study of fundamental optical properties of $MoS_2$ revealed by optical spectroscopy of Raman, photoluminescence, and vacuum ultraviolet spectroscopic ellipsometry. A band gap of 1.42 eV is determined by the absorption threshold of bulk $MoS_2$ that shifts to 1.83 eV in monolayer $MoS_2$. We extracted the high precision dielectric function up to 9.0 eV which leads to the identification of many unique interband transitions at high symmetry points in the $MoS_2$ momentum space. The positions of the so-called A and B excitons in single layers are found to shift upwards in energy compared with those of the bulk form and have smaller separation because of the decreased interactions between the layers. A very strong optical critical point predicted to correspond to a quasi-particle gap is observed at 2.86 eV, which is attributed to optical transitions along the parallel bands between the M and Γ points in the reduced Brillouin zone. The absence of the bulk $MoS_2$ spin-orbit interaction peak at ~ 3.0 eV in monolayer $MoS_2$ is, as predicted, the consequence of the coalescence of nearby excitons. A higher energy optical transition at 3.98 eV, commonly occurred in bulk semiconductors, is associated with a combination of several critical points. Additionally, extending into vacuum ultraviolet energy spectrum are a series of newly observed oscillations representing optical transitions from valence bands to higher conduction bands of monolayer $MoS_2$ complex band structure. These optical transitions herein reported enhance our understanding of monolayer $MoS_2$ as well as of two-dimensional systems in general, and thus provide informative guidelines for $MoS_2$ optical device designs and theoretical considerations.

# I. Introduction

MoS$_2$, an emerging new class of atomically thin material down to a unit cell thickness (monolayer), has shown intriguing physical properties and exciting prospects for a variety of applications.[1-7] The monolayer MoS$_2$ is composed of a hexagonal plane of Mo atoms sandwiched between two hexagonal planes of S atoms in a trigonal prismatic arrangement, while its bulk counterpart can be considered as a stack of many S-Mo-S sheets weakly bonded by van der Waals force.[8] Unlike graphene, there is an intrinsic band gap in MoS$_2$, which changes from an indirect gap of E$_g \sim$ 1.3 eV in bulk to a direct gap in monolayer.[8, 9] Therefore, monolayer MoS$_2$ is more attractive than graphene for future transistors and logic circuit applications where a high on/off current ratio is required.[2, 5, 10-12] More importantly, due to the fact that the band gap of monolayer MoS$_2$ is direct with the energy in the visible range of the spectrum, it could be used in light emitting diodes, photodetectors or solar cells, and transparent, light, and flexible devices.[1, 2, 4, 6, 13] With such a broad potential application base, a comprehension of the optical properties of MoS$_2$ will elucidate its electronic band structure which is critical to electronic and optoelectronic devices researchers. It is known that the connection of the microscopic details manifested in the electronic band structure of a solid with experimental macroscopic observables can be established through a response function. In the case of probing photons and interacting electrons, the commonly used response function is the wave vector and frequency dependent complex dielectric function.[14] In most previous experimental studies of monolayer or ultra-thin MoS$_2$, small flakes (several to tens of micrometers) were produced by mechanical exfoliation[3, 5, 8, 9] Such small sample size generally hinders the measurement of the complex dielectric function of MoS$_2$ by optical techniques. An up-to-date literature survey shows the optical response of MoS$_2$ has been reported by only a few research groups and most are theoretical studies.[8, 9, 15-18] The reported experimental studies were focused on photoluminescence (PL), absorbance, and photo conductivity in the vicinity of the K point in its Brillouin zone and the corresponding spectral ranges are rather limited.[8, 9, 19-21] On the other hand, the broad band optical dielectric function of monolayer $MoS_2$ has not been reported and it is expected to be distinct from that of bulk $MoS_2$. Not until very recently could large area monolayer $MoS_2$ with sample size up to millimeters be grown by CVD.[22-25] Such a large size sample facilitates the optical measurement of $MoS_2$ optical response function.

In this report, a comparative study of broad band (1 eV to 9 eV) dielectric function of monolayer and bulk $MoS_2$ is experimentally reported for the first time. The optical band gap of 1.42 eV was observed for bulk $MoS_2$, but shifts to a higher energy of 1.83 eV for single layer $MoS_2$. The so-called A and B excitons due to the direct d-d transitions in monolayer are accurately derived to be 1.88 eV and 2.02 eV by line-shape fitting of the experimental data, which are blue-shifted as compared with those of the bulk form. A series of optical critical points (CPs) at higher energy than the A and B excitons are clearly observed with a very strong absorption at 2.86 eV, which is attributed to optical transitions along the parallel bands between the M and Γ points in the reduced Brillouin zone. Above these transitions, a critical point at 3.98 eV is observed and can be associated with a combination of several critical points as seen for common semiconductors.[14] Deep in the ultraviolet (UV) region, the dielectric function shows a series of absorption peaks or fine features separated by about 0.7 eV.

## II. Experimental Section

Chemical vapor deposition (CVD) in a hot-wall furnace is adopted to synthesize a monolayer of $MoS_2$ directly on $SiO_2$/Si substrates using $MoO_3$ and S powders as the reactants. The substrates are treated to make the surface hydrophilic prior to the application of PTAS (perylene-3,4,9,10-tetracarboxylic acid tetrapotassium salt). As discussed in reference 26, the PTAS promotes monolayer growth of $MoS_2$ by serving as a nucleation site for the precursors. Large-area $MoS_2$ layers can be directly obtained on amorphous $SiO_2$ surfaces without the need to use highly crystalline metal substrates or an ultrahigh vacuum environment. The resulting $MoS_2$ films are highly crystalline and their size is up to several millimeters. Samples were transferred from the growth substrate using the PMMA mediated wet method. PMMA was spin-coated on the samples and allowed to dry at room temperature overnight. The PMMA-supported MoS$_2$ was released from the Si/SiO$_2$ substrate in a 0.2 M KOH solution heated to 60ºC. The sample was cleaned in three DI water baths and then placed on the fused silica substrate. After the sample was allowed to dry at 40ºC, it was baked at 180ºC to improve adhesion between the MoS$_2$ and the substrate. The PMMA was removed via an overnight soak in acetone. Bulk MoS$_2$ was commercially purchased from SPI Supplies[*].

Spectroscopic ellipsometry measurements were performed on a vacuum ultraviolet spectroscopic ellipsometer with a light photon energy from 1.0 eV to 9.0 eV in steps of 0.01 eV. The SE data were taken at multiple angles of incidence of 65°, 70°, and 75°. The three-phase ellipsometric model consisting of the substrate (SiO$_2$), thin film (MoS$_2$), and ambient (air) was employed to extract the dielectric function of MoS$_2$. Without explicitly expressing the ellipsometric equation of the 3-phase structure, it can be written as:

$$\tan \psi(\lambda) \, exp(i\Delta(\lambda)) = f\left(\varepsilon_{SiO_2}, \varepsilon_{MoS_2}, \varepsilon_{Air}, d_{MoS_2}, \phi, \lambda, \right) \qquad (3)$$

The known parameters in this equations are the dielectric function of SiO$_2$ ($\varepsilon_{SiO_2}$) and air ($\varepsilon_{Air}$), the thickness of MoS$_2$ ($d_{MoS_2}$), the angle of incidence ($\phi$), and wavelength ($\lambda$). The only two unknown parameters are the real and imaginary parts of the complex dielectric function of MoS$_2$ ($\varepsilon_{MoS_2}$). Therefore, with two measurable parameters $\psi$ and $\Delta$, $\varepsilon_1$ and $\varepsilon_2$ of MoS$_2$ can be uniquely calculated. Micro-Raman and PL measurements were performed with a WITec Alpha 300RA[*] system using the 532 nm line of a frequency-doubled Nd:YAG laser as the excitation source. The spectra were measured in the backscattering configuration using a 100 × objective and either a 600 or 1800 grooves/mm grating. The spot size of the laser was ~ 342 nm resulting in an incident laser power density of ~ 140 µW/µm$^2$. A low laser power was used in order to avoid any significant heat related effects in the Raman or PL signatures.[27] Raman and PL maps were developed over 25 μm × 25 μm regions using a 250 nm grid spacing. The integration time was 5 seconds for Raman and 1 second for PL.

---

[*] Certain commercial equipment, instruments, or materials are identified in this report in order to specify the experimental procedure adequately. Such information is not intended to imply recommendation or endorsement by the National Institute of Standards and Technology, nor is it intended to imply that the materials or equipment identified are necessarily the best available for the purpose.

### III. Results and discussion

Fig. 1(a) shows the optical image of the transferred sample, which appears to be free from contaminations and uniform (also see Fig. S1). Raman spectroscopy is commonly used to characterize the thickness of $MoS_2$. Since there is no center of inversion for one molecular layer of $MoS_2$, $E'$ and $A'$ are used instead of the common $E^1_{2g}$ and $A_{1g}$ notations. These two characteristic Raman modes, $E'$ at ~ 385 $cm^{-1}$ and $A'$ at ~ 405 $cm^{-1}$, are sensitive to $MoS_2$ layer thickness. As the number of $MoS_2$ layers increases, the frequency of $E'$ decreases, while the frequency of $A'$ increases.[28, 29] Thus, the frequency difference between $E'$ and $A'$ is often used to identify the number of $MoS_2$ layers. Fig. 1(b) shows the Raman spectra averaged over the 5 μm × 5 μm center area (400 spectra) of the sample in Fig. 1(a). The wave number difference between the $E'$ and $A'$ peak is 19 $cm^{-1}$, thus confirming that our $MoS_2$ sample is monolayer.[28] Fig. 1(c) and (d) display the Raman intensity mapping images of the $E'$ and $A'$ peak, respectively, which were taken from the whole area shown in Fig. 1(a) confirming the uniformity of a monolayer across the area.

It is known that as the thickness of $MoS_2$ decreases from bulk to a monolayer, the band gap crosses over from an indirect to direct gap.[8] Thus, the PL quantum efficiency of monolayer $MoS_2$ enhances significantly, more than four orders of magnitude compared with the bulk material.[8] PL spectra were also obtained for our CVD-grown monolayer $MoS_2$ samples. Shown in Fig. 2(a) is the PL spectra averaged over the 5 μm × 5 μm center area (400 spectra) pictured in Fig. 1(a). Two PL peaks, known as A and B excitons,[8, 16] are observed at ~ 1.86 eV and ~ 2.00 eV, respectively, which are consistent with a previous report.[9] Shown in Fig. 2(b) and 2(c) are the PL mapping images of the A exciton peak intensity and energy, respectively. These maps were taken on the whole area of Fig. 1(a), which revealed slightly more variations, up to 50 meV in energy (min. to max.) across the sample, than the uniform Raman maps as shown in Fig. 1(c) and (d). Such variations may be due to the structural defects in the sample, which are difficult to detect by the E′ and A′ Raman peaks intensity mapping, but more easily by the PL energy and intensity mapping.[20] Another possible reason for the variations is the result of the fluctuating nature of the interfacial contact between the substrate and MoS$_2$ that might affect the dielectric screening of the long-range Coulomb interaction.[1, 19]

The broadband dielectric function is measured by Vacuum Ultraviolet Spectroscopic Ellipsometry (SE) in this study, as schematically illustrated in Fig. 3(a). SE is a non-destructive and non-contact technique widely used to characterize optical properties of thin films.[30] Compared to other optical techniques such as transmission and reflection spectroscopy, absorbance and photo conductivity, SE is more advantageous in that both the real ($\varepsilon_1$) and imaginary ($\varepsilon_2$) part of the complex dielectric function are obtained simultaneously. Once $\varepsilon_1$ and $\varepsilon_2$ are known, other optical parameters of the material, e.g., refractive index, absorption coefficient, reflectivity and loss function can be easily derived. The large area and high quality of MoS$_2$ films can easily accommodate the ellipsometry light beam size and yield very low noise signals in very wide energy range (1~9 eV), from which high order numerical differentiations of the dielectric function can be carried out with high precision. Shown in Fig. 3(b) and (c) are the extracted real ($\varepsilon_1$) and imaginary parts ($\varepsilon_2$), respectively, of the monolayer MoS$_2$. For comparison, the dielectric function of the bulk material was also measured with SE. For monolayer MoS$_2$, the most prominent features in the spectra are a series of sharp peaks, i.e., CPs, in the lower energy range, and a series of broad and low oscillations at higher photon energies. The sharp features in $\varepsilon_2$ are classified according to a standard nomenclature, where the lowest energy structural feature is denoted by $E_0$ corresponding to the direct-gap transitions from the valence band maximum to the conduction minimum band at K-point in the Brillouin zone.[14] $E_0$ is immediately followed by the $E_0+\Delta_0$ peak, which corresponds to the spin-orbital splitting of the valence band at the same K-point. These two features are designated as A and B excitons by PL measurements, and they are attributed to be direct-gap transitions.[8, 9, 31] There appears to be a slight shift in energy of the peak $E_0$ and $E_0+\Delta_0$ when compared with peaks A and B obtained by PL shown in Fig. 2(a). In fact, there is a slight variation of about 50 meV of the A peak position across the sample as shown in Fig. 2, which is larger than the shift mentioned above. As a result, within the variation, it can be said that the energies of A and B peaks are the same as those of the $E_0$ and $E_0+\Delta_0$ CPs. It is interesting to note that, in contrast to the PL spectra reported by other groups[8, 20] where the B peak was absent from monolayer $MoS_2$ but present for thicker layers, our monolayer PL clearly shows the B peak with a lower luminescence intensity than A peak, which was similarly observed in Ref. 9.

It is just not accurate to extract the energy critical point by simply locating the energy at which the maximum of absorption spectrum occurs.[32] To accurately determine the energy position of the CPs, it is necessary to locally isolate a CP from the polarization contributions from other nearby CPs. One direct experimental approach is to apply an external perturbation as modulations such as electric or magnetic field or stress to suppress uninteresting background effects and strongly reveal structures localized in the energy.[14] Another approach is to produce a similar modulation response by numerical differentiation of the dielectric function that in effect filters out the above-mentioned background.[14] In the vicinity of a CP, the dielectric response can be represented by the following analytic form:[33, 34]

$$\varepsilon(E) = C + A\Gamma^{-n}e^{i\theta}(E - E_t + i\Gamma)^n, \qquad (1)$$

where each CP is described by five energy independent parameters: the amplitude A, phase angle θ, threshold energy $E_t$, and phenomenological broadening Γ which relates to scattering rates, and *n*. The exponent *n* characterizes the analytical shape of ε(E) near its minima or maxima and depends on the signs of the reduced masses of the electron and hole. The value of n is -1/2 for one-dimensional (1D), 0 (logarithmic: $\ln(E - E_t + i\Gamma)$ ) for two-dimensional (2D), and +1/2 for three-dimensional (3D) CP. The critical point dimension is classified based on the electron energy dispersion at Van Hove singularities which can be expanded as a function of momentum vector k about the critical point as $E(k) \sim E(0) + \alpha_1 k_1^2 + \alpha_2 k_2^2$

+ $\alpha_3 k_3^2$ + … The dimension is assigned by the number of non-zero coefficients $\alpha$'s. 1D, 2D and 3D represent one, two, or three, respectively, non-zero $\alpha$'s.[14, 33] The direct discrete bound exciton is included in the same equation for which $n = -1$. Since the measured $\varepsilon(E)$ is a superposition of different spectra calculated from equation (1) and in order to eliminate the effect of the constant background, we perform line-shape fitting to the second derivative of the dielectric function with respect to photon energy:

$$\frac{d^2\varepsilon}{dE^2} = \begin{cases} n(n-1)A\Gamma^{-n}e^{i\theta}(E - E_t + i\Gamma)^{n-2}, & n \neq 0 \\ -A\Gamma^{-n}e^{i\theta}(E - E_t + i\Gamma)^{-2}, & n = 0 \end{cases} \quad (2)$$

To ensure an accurate determination of $E_t$, the CP energy, the fitting was carried out on the real and imaginary parts of $\frac{d^2\varepsilon}{dE^2}$ simultaneously. Shown in Fig. 4(a) and (b) are the fitting of equation (2) to the second derivative of the measured dielectric function of the monolayer $MoS_2$.

For the $E_0$ and $E_0+\Delta_0$ peaks, it was found that $n = -1$ yields the best fit, which explicitly confirms that they are excitons. First-principle calculations have predicted that the exciton peak at K point in the Brillouin zone is split into two peaks at ~1.88 eV and ~2.02 eV, respectively, by spin-orbital coupling.[16] These two peaks around those energies are observed experimentally and ascribed to the A and B excitons as we discussed above. The insets of Fig. 4(a) and (b) are intended to show that the best fit is obtained by using a discrete exciton when compared with the 1D (not shown), 2D (not shown), and 3D (shown as blue lines by the inset of Fig. 3(c)) line shape. The effect of the electron-hole interaction, i.e., excitonic effects, is to enhance the oscillator strength of the transitions that result from the interaction of the discrete exciton with a continuous background.[33, 34] Without the excitonic effects the theory reproduces the low-energy exponential absorption edge of $\varepsilon_2$ but fails to account for the strength of the negative and positive peak. By including Coulomb effects, the discrepancy is eliminated and a good fit is obtained. As a fitting result of equation (2) by using n = -1, the fitted peak positions are 1.88 eV and 2.02 eV for the $E_0$ and $E_0+\Delta_0$ peaks, respectively. Table 1 is a compilation of recent theoretical and experimental results showing SE $E_0$ and $E_0+\Delta_0$ peaks agree with the latest first-principles calculations.[16] Compared with bulk $MoS_2$, where the separation of the A and B peak is about 140 meV, these two peaks in monolayer blue shift 50 meV and 10 meV, respectively (see Fig. 3(b)), which leads to a narrower separation. Such blue shift and smaller separation in a monolayer result from the decreased interactions between the layers.[8, 35]

Table 1. The optical transition energies of peaks A and B and quasiparticle band gap from different studies: $G_1W_0$, $G_0W_0$-BSE, and sc-$GW_0$-BSE are theoretical models all using GW-Bethe-Salpeter equation but with different physical parameters (see Ref. 16).

|  |  | This work (SE) | $G_1W_0$ [16] | Absorp. [8] | PL [9] | $G_0W_0$-BSE [17] | sc-$GW_0$-BSE[18] |
|---|---|---|---|---|---|---|---|
|  |  | Experiment | Theory | Experiment | Experiment | Theory | Theory |
| Optical transition energies (eV) | A | 1.88 | 1.88 | 1.88 | 1.83 | 1.78 | 2.22 |
|  | B | 2.02 | 2.02 | 2.03 [§] | 1.98 | 1.96 | 2.22 |
| Quasiparticle band gap (eV) |  | 2.86 | 2.84 | 2.86 [¶] | ... | 2.82 ($G_0W_0$) | 2.80 (sc-$GW_0$) |

[§],[¶] Extracted from figure 4(a) and figure S1(a), respectively, of Ref. 8

The next higher energy feature in the $\varepsilon$ spectra is denoted by $E_1$. For the monolayer $MoS_2$, it is found that the line-shape fitting to this critical point by equation (2) yields the same figure of merit for different values of *n* and yields a resonance of 2.86 eV (see Fig. 4). In contrast to monolayer $MoS_2$, a lower or red shifted $E_1$ of ~ 2.60 eV is observed for bulk $MoS_2$. From the first principle full-potential linearized augmented plane wave (FLAPW) band structure calculation by Yun et al.,[36] they show that the lowest conduction band and the highest valence band are parallel over a wide range in the Brillouin zone between M and Γ points, thus leading to a maximum in the joint density of states, and therefore giving rise to the critical point $E_1$. The energy of $E_1$ for the bulk $MoS_2$ is consistent with the first-principle calculated value of 2.60 eV along the M-Γ direction.[36] An additional peak labeled as $E_1+\Delta_1$ at ~ 3.0 eV is observed in bulk $MoS_2$, which is due to the large spin-orbit interaction in the valence band, but absent in the spectrum of monolayer.[37] For the monolayer, it is interesting to note that the fitted resonant energy at 2.86 eV ($E_1$) matches quite well with the theoretical quasi-particle gap (~ 2.84 eV) at K point.[16] It has been also theoretically predicted that electron-phonon interaction in monolayer $MoS_2$ will produce few excitons in the energy range from 2 eV to 3 eV. These excitons are expected to coalesce into a much broader peak.[16] In other words, it might be reasonable to speculate that the $E_1+\Delta_1$ peak observed in bulk, but absent in monolayer, converges into $E_1$ peak of monolayer. Also, it might explain that all line shape dimensions (1D, 2D, 3D, and exciton) have the same goodness-of-fit to the peak at $E_1$ as discussed above (see Fig. 4).

Above $E_1$ is the $E_2$ critical point whose energy resonance position obtained from the fitting of equation (2) to monolayer $MoS_2$ yields a value of 3.98 eV (see Fig. 4(a) and 4(b)). CPs at this energy range or above were seldom reported experimentally, and the origin of the $E_2$ edge is not well identified. We speculate it is associated with a combination of several critical points next to each other similarly observed for common semiconductors.[14]

Above $E_2$ critical points, a series of broad and low oscillations are observed in the $\varepsilon_2$ spectra. These oscillations, with an energy separation of about 0.7 eV, can be seen more clearly by the corresponding absorption coefficient, as shown in the inset of Fig. 3(c). These oscillations are actual responses from the $MoS_2$ film, not artifacts from the silica substrate or the possible interferences with the backside of the substrate (see Fig. S2). They are the results of optical transitions from valence bands to higher conduction bands and are difficult to identify with other higher critical points in its complex band structure.[36] To the best of our knowledge, these high energy fine structures were neither discussed in theory nor reported in experiments before. Therefore, further efforts are needed to elucidate these oscillations, as well as the origin of the $E_2$ critical point.

For direct gap semiconductors, $\varepsilon_2$ can be expressed as $\varepsilon_2 = C\left(\frac{E}{E_g}\right)^{-2}\left(\frac{E}{E_g}-1\right)^{1/2}$ near the direct energy gap ($E_g$), where C is a constant related to the transition probability from the valence band to conduction band and E is the photon energy.[14] The expression can be rewritten as $\varepsilon_2^2 E^4 \propto E - E_g$, and therefore the plot of $\varepsilon_2^2 E^4$ as a function of photon energy E yields the optical band gap value, $E_g$, highly precisely. This is illustrated by the red curve in Fig. 5 from which the optical bandgap of the monolayer $MoS_2$ is extracted to be 1.83 eV. To obtain the band gap or absorption edge of the bulk $MoS_2$ which is an indirect gap semiconductor, the SE transmission measurement was performed instead of the ellipsometry reflectivity measurement, because the surface of our bulk sample is not quite specular on a large surface scale. The ellipsometric parameter, $\tan\psi$ shown in Fig. 5, is proportional to the reflectivity[30] of the light incidence on bulk $MoS_2$ yielding a band gap of 1.42 eV which is within the range of values reported in literature as an indirect band gap.[8, 38]

## IV. Conclusions

In summary, the optical properties of CVD-grown large area monolayer $MoS_2$ are presented in this report. Raman and photoluminescence spectroscopy studies proved our $MoS_2$ samples are monolayer with high crystalline quality and uniformity. Strong photoluminescence peaks are observed due to the well-known A and B excitonic transition. The broad band optical dielectric function, from 1 eV to 9 eV, is measured by the extended spectral range spectroscopic ellipsometry for the first time. In addition to the $E_0$ and $E_0 + \Delta_0$ critical points corresponding to A and B photoluminescence peaks, a series of higher energy sharp peaks, $E_1$, $E_1 + \Delta_1$ and $E_2$, are observed in the dielectric function of $MoS_2$. The energies of these CPs were precisely obtained by a line shape fitting method, and the results agree with theoretical calculation very well. Also, a series of absorption peaks separated by ~ 0.7 eV are clearly detected at higher UV photon energies. We believe that the optical features first reported here will stimulate further investigations, especially to shed light on their origins from the band structure calculation perspectives and, in addition, will provide better understanding of the optical responses of MoS$_2$ to their optoelectronic device designs and applications.

**Acknowledgement**:

This work was supported by the Ministry of Science and Technology of China (Grant No. 2011CB921904) and the Ministry of education of China (Grant No. 113003A). W. L. was partly supported by the National Institute of Standards and Technology. J. K. acknowledges the support through the STC Center for Integrated Quantum Materials from NSF grant DMR-1231319. A. G. B., M. A. and R. A. B. were supported by the U.S. Army Research Laboratory (ARL) Director's Strategic Initiative program on interfaces in stacked 2D atomic layered materials.

**Figure captions**

**Figure 1**. Raman characterization of the CVD-grown $MoS_2$ on fused silica using a 532 nm laser line with a 342 nm spot size and 12.5 μW power: (a) Optical image of the monolayer $MoS_2$ on fused silica substrate. (b) Averaged Raman spectroscopy of a 5 μm × 5 μm area in the center of the area shown in (a). (c) and (d) Raman mapping images of the $E'$ and $A'$ peak intensity, respectively, taken in the whole area shown in (a) with a 250 nm grid spacing. Scale bar: 5 μm.

**Figure 2**. PL characterization of the CVD-grown $MoS_2$. (a) Averaged PL spectrum (red curve) measured over 5 × 5 μm area in the center of the area in Fig. 1(a) and the imaginary part ($\varepsilon_2$) (blue curve) of the dielectric function measured by spectroscopic ellipsometry. (b) and (c) are PL mapping images of the A exciton peak intensity and peak energy, respectively, taken in the area shown in Fig. 1(a) with a 250 nm grid spacing. Scale bar: 5 μm.

**Figure 3**. The schematic of ellipsometry measurement and the dielectric function of the CVD-grown monolayer $MoS_2$. (a) Schematic of SE measurement. (b) The real ($\varepsilon_1$) part of the dielectric function of CVD-grown monolayer and bulk $MoS_2$, inset: zoom-in of the spectra range from 1.6 eV to 3.2 eV. (c) The imaginary ($\varepsilon_2$) part of the dielectric function of CVD-grown monolayer and bulk $MoS_2$. The inset is the absorption spectrum of monolayer $MoS_2$ calculated from the measured dielectric function.

**Figure 4**. Critical point determination. Red curves are the fits to the second derivatives (black curves) of the real ($\varepsilon_1$) and imaginary ($\varepsilon_2$) parts of the dielectric function of monolayer $MoS_2$. The inset is the comparison of the fitting results when using exciton and 3D lineshapes.

**Figure 5.** Determination and comparison of the optical band gap of monolayer (blue curve) and bulk (red curve) $MoS_2$.

Figure 1

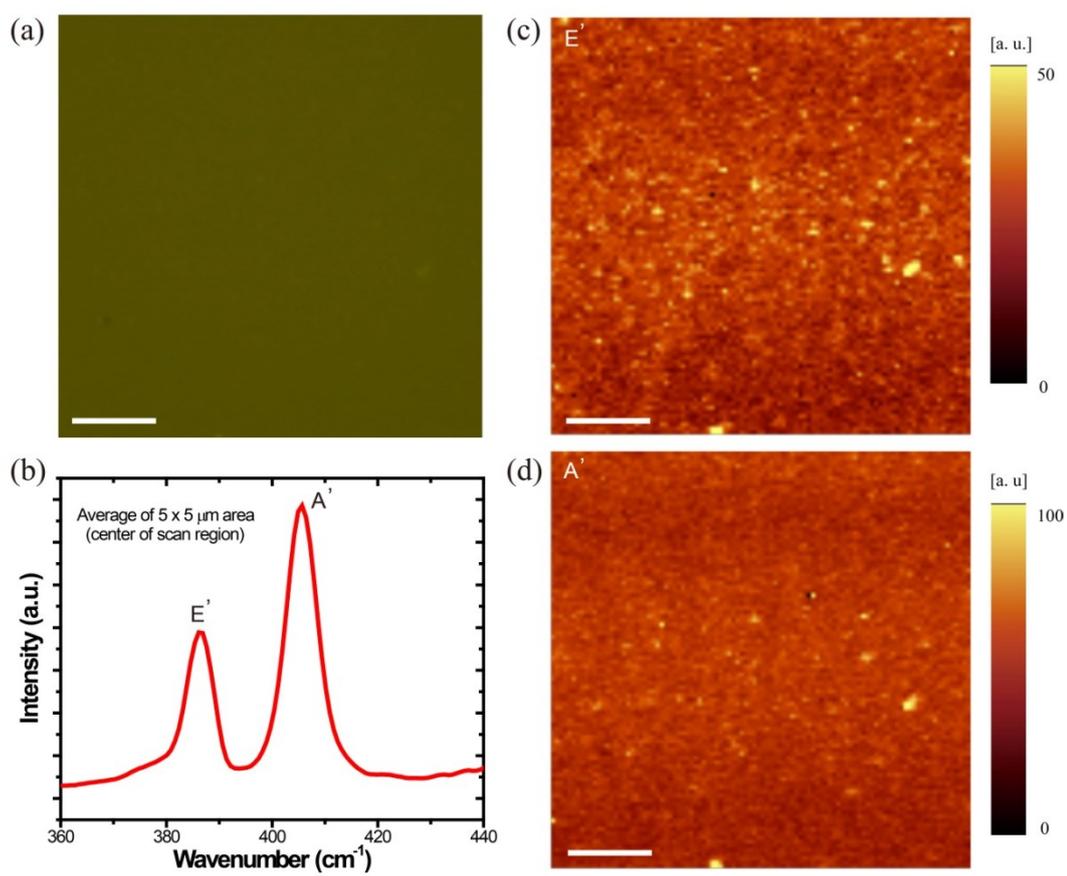

Figure 2

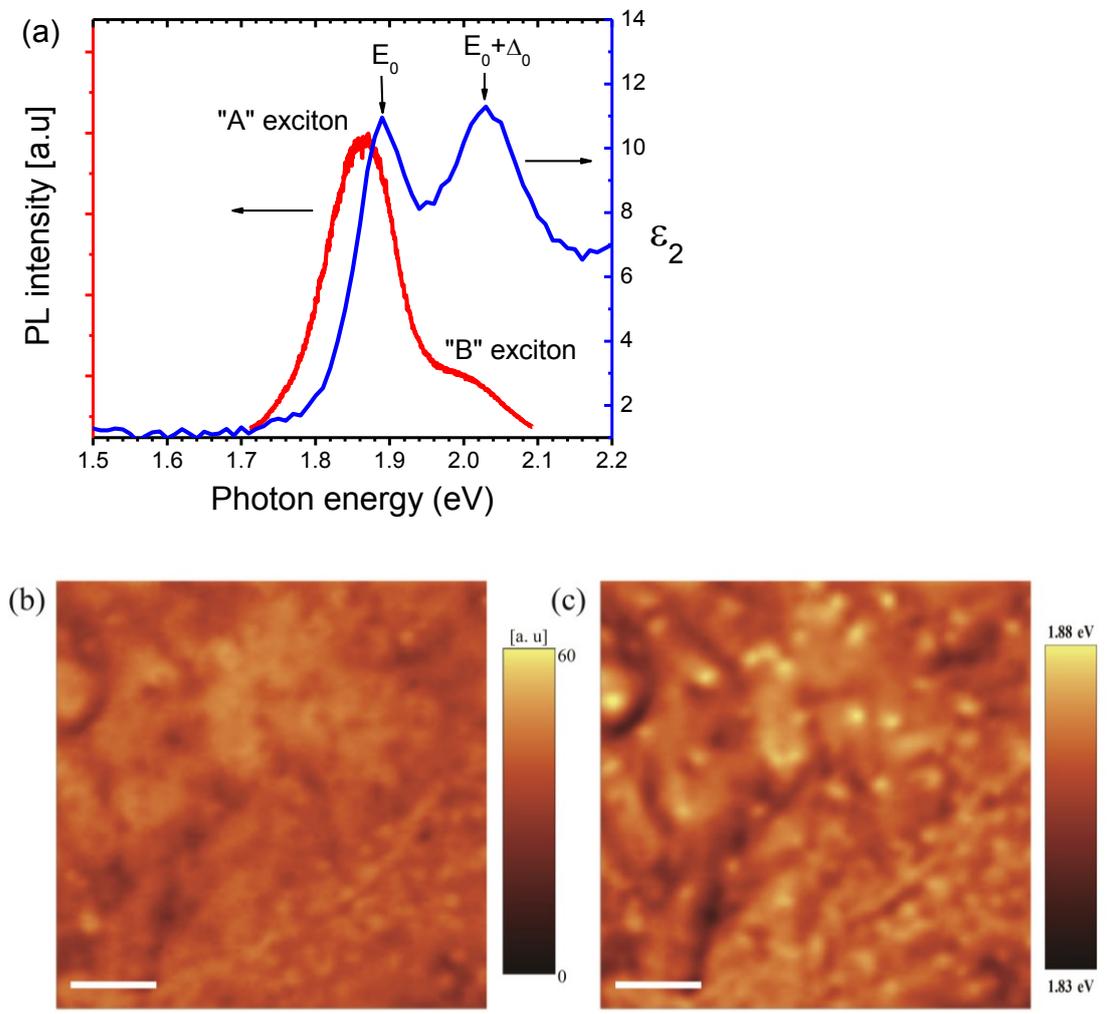

Figure 3.

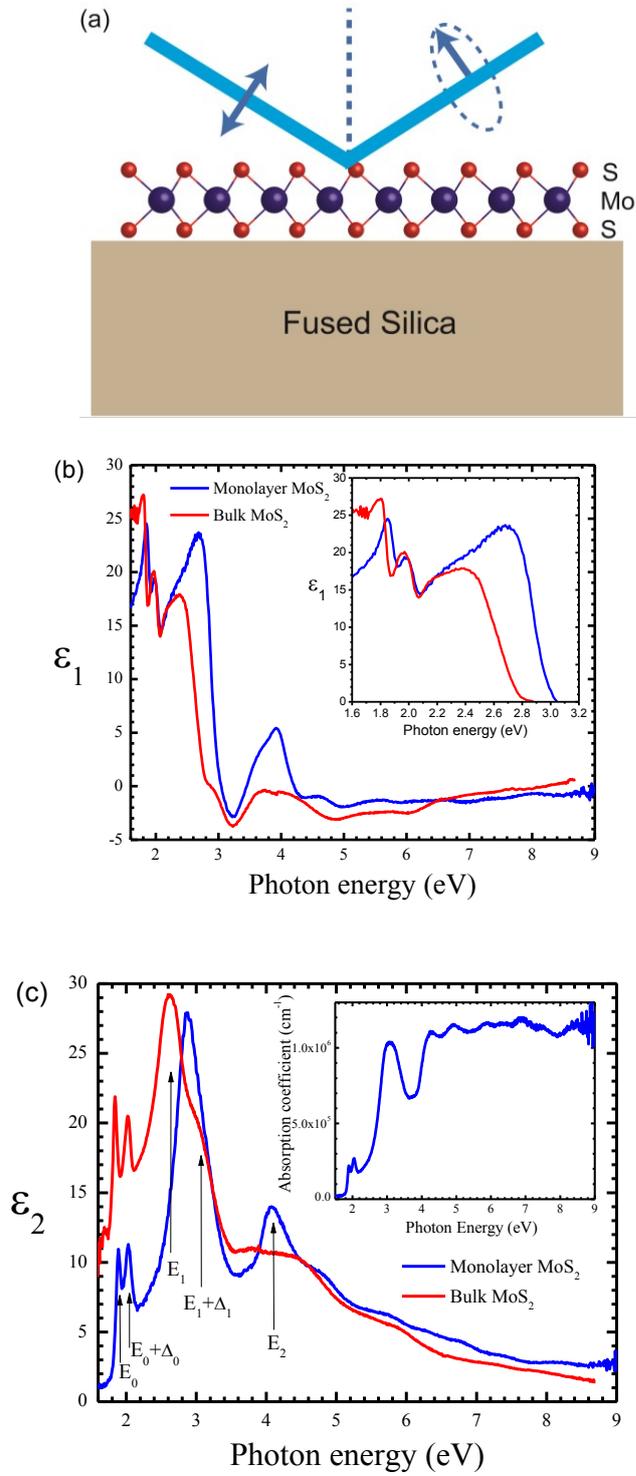

Figure 4:

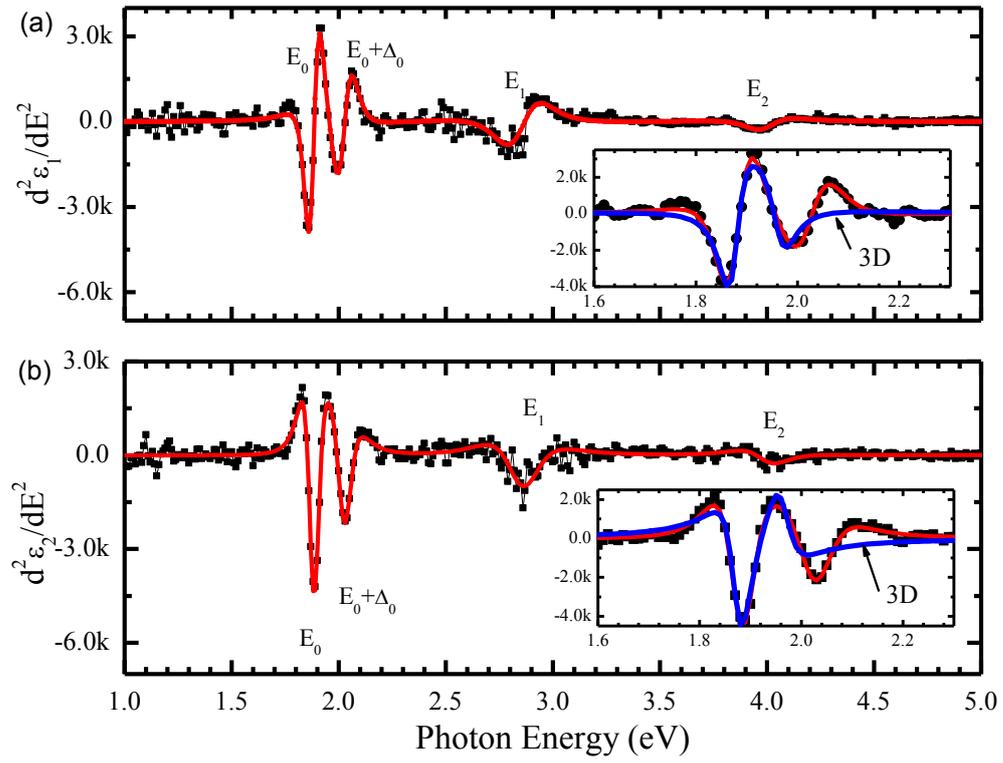

Figure 5:

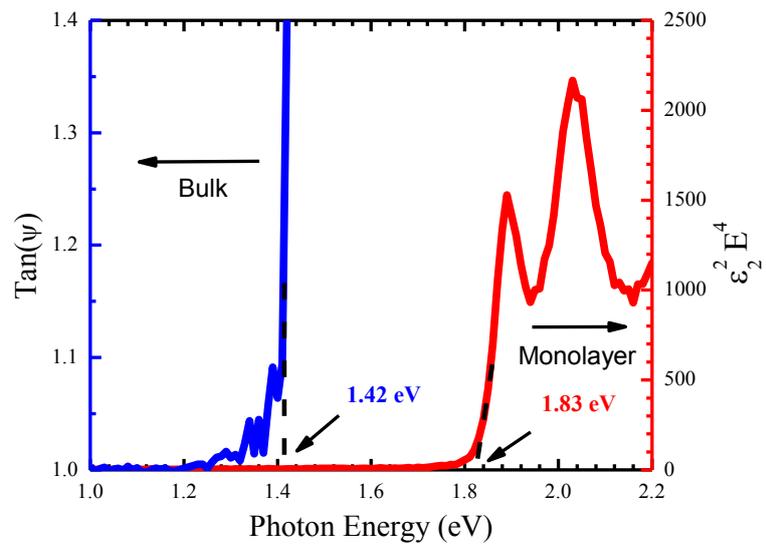